      [1999/12/01 v1.4c Il Nuovo Cimento]
\documentclass{cimento}

\usepackage{graphicx}  
\usepackage{bm}
\newcommand{\be}{\begin{equation}}
\newcommand{\ee}{\end{equation}}
\newcommand{\bea}{\begin{eqnarray}}
\newcommand{\eea}{\end{eqnarray}}
\newcommand{\bfp}{\bm{p}}

\newcommand{\qup}{q^\uparrow}

\newcommand{\pup}{p^\uparrow}
\title{On the potential role of the Collins effect in $A_N$ in $pp\to\pi\,X$}
\author{M.~Anselmino\from{ins:a}\ETC, 
M.~Boglione\from{ins:a}, 
U.~D'Alesio\from{ins:c}\from{ins:d}\thanks{Presenting author; talk delivered at the ``3rd Workshop on the QCD Structure of the Nucleon" (QCD-N'12), Oct. 22-26, 2012, Bilbao, Spain.},
E.~Leader\from{ins:e},
S.~Melis\from{ins:f},
F.~Murgia\from{ins:d}
\atque
A.~Prokudin\from{ins:g}}
\instlist{\inst{ins:a} Dipartimento di Fisica Teorica, Universit\`a di Torino, and INFN, Sezione di Torino, Italy
\inst{ins:c} Dipartimento di Fisica, Universit\`a di Cagliari, Monserrato (CA), Italy
\inst{ins:d} INFN, Sezione di Cagliari, C.P.~170, Monserrato (CA), Italy
\inst{ins:e} Imperial College London, South Kensington Campus, United Kingdom
\inst{ins:f} Eur.~Centre for Theor. Studies in Nucl. Physics and Related Areas (ECT*), Trento,
Italy
\inst{ins:g} Jefferson Laboratory, Newport News, Virginia 23606, USA}
\PACSes{\PACSit{13.88.+e, 12.38.Bx, 13.85.Ni}{}
}
\begin{document}

\maketitle

\begin{abstract}
  Transverse single spin asymmetries in $p p \to \pi\, X$ processes, while on a quite firm ground experimentally, are still a much debated phenomenological issue. We consider them in a transverse momentum dependent factorization scheme. After revisiting a previous result, we give new estimates of the Collins contribution by adopting the latest information on the Collins and transversity functions, as extracted from SIDIS and $e^+e^-$ data.
\end{abstract}

\section{Introduction}

Transverse single spin asymmetries (SSAs), observed in inclusive and semi-inclusive processes, show amazing features and still represent a challenging topic in spin physics.
We consider here the SSAs for inclusive pion production in $pp$ collisions within a transverse momentum dependent (TMD) factorization scheme (see, e.g., Ref.~\cite{D'Alesio:2007jt} for a review). For such SSAs there are many possible contributions, as shown in Refs.~\cite{Anselmino:2004ky,Anselmino:2005sh}, where it was argued that only the Sivers effect could be sizeable, being all the other contributions strongly suppressed by phase integrations. Unfortunately, that conclusion was affected by a wrong sign in one of the elementary interactions involving the Collins effect that was therefore underestimated.
On this basis we revisit the Collins pion SSA, taking into account the phenomenological information so far extracted from data on azimuthal asymmetries in semi-inclusive deep inelastic scattering (SIDIS) and $e^+e^-$ processes.

\section{The role of the Collins effect in $A_N$}
The SSA for the process $\pup p\to \pi\, X$ is defined as:
\be
A_N = \frac{d\sigma^\uparrow - d\sigma^\downarrow}
           {d\sigma^\uparrow + d\sigma^\downarrow} \label{an}
\quad\quad {\rm where} \quad\quad
d\sigma^{\uparrow, \downarrow} \equiv
\frac{E_\pi \, d\sigma^{p^{\uparrow, \downarrow} \, p \to \pi \, X}}
{d^{3} \bfp_\pi} \>\cdot
\ee
In a TMD approach (adopting a schematic form) the Collins contribution reads~\cite{Anselmino:2004ky,Anselmino:2005sh}:
\bea
\mbox{}\;\;\;\;[d\sigma^\uparrow - d\sigma^\downarrow]_{\rm Collins}
&=& \sum_{q_a,b,q_c,d}\!\!
\Delta_Tq_a(x_a, k_{\perp a}) \otimes f_{b/p}(x_b, k_{\perp b}) \otimes
d\Delta\hat\sigma \otimes
\Delta^N\! D_{\pi/\qup_c}(z, p_\perp) \>,
\eea
showing the coupling of the transversity, $\Delta_Tq_a$, with the Collins fragmentation
function, $\Delta^N\! D_{\pi/\qup_c}$ \cite{Collins:1992kk}. The quantity $d\Delta\hat\sigma$ is the elementary spin transfer for the process $q_a b \to q_c d$.
After correcting a sign mistake in $d\Delta\hat\sigma$ for the $q g \to q g$ channel, we can now give new realistic estimates of the contribution of the Collins effect to $A_N$ (see Ref.~\cite{Anselmino:2012rq} for details).

Let us start recalling the main aspects of the analysis of SIDIS and $e^+e^-$ data. We adopt a simple factorized form of the TMD functions:
\begin{equation}
f_{q/p}(x,k_\perp) = f_{q/p}(x)\,
\frac{e^{-k_\perp^2/\langle k_\perp^2 \rangle}}
{\pi \langle k_\perp^2 \rangle}\,,
\label{eq:pdf-unp}
\quad\quad\quad
D_{\pi/q}(z,p_\perp) = D_{\pi/q}(z)\,
\frac{e^{-p_\perp^2/\langle p_\perp^2 \rangle}}
{\pi \langle p_\perp^2 \rangle}\,,
\label{eq:ff-unp}
\end{equation}
where $\langle k_\perp^2\rangle = 0.25\, {\rm GeV}^2$ and $\langle p_\perp^2\rangle = 0.20\, {\rm GeV}^2$ \cite{Anselmino:2005ea}. For the usual $k_\perp$-integrated $f_{q/p}(x)$ we use the GRV98
set~\cite{Gluck:1998xa} and for the $p_\perp$-integrated
$D_{\pi/q}(z)$ the Kretzer set~\cite{Kretzer:2000yf}. 
The quark transversity distribution, $\Delta_T q(x,k_\perp)$, and the Collins fragmentation function (FF), $\Delta^N\! D_{\pi/q^\uparrow}(z,p_\perp)$, are parameterized as follows:
\bea
\Delta_T q(x,k_\perp) &=& \frac{1}{2}\,{\cal N}_q^T(x)\,\left[\,f_{q/p}(x)+
\Delta q(x)\,\right]\,\frac{e^{-k_\perp^2/\langle k_\perp^2 \rangle_T}}
{\pi \langle k_\perp^2 \rangle_T}\label{detq}\\
\Delta^N\! D_{\pi/q^\uparrow}(z,p_\perp) &=& 2 {\cal N}_q^C(z)\,D_{\pi/q}(z)\,
\sqrt{2e}\,\frac{p_\perp}{M_h}\,e^{-p_\perp^2/M_h^2}\,\frac{e^{-p_\perp^2/\langle p_\perp^2 \rangle}}
{\pi \langle p_\perp^2 \rangle}\,,
\eea
where $\Delta q(x)$ is the usual collinear quark helicity distribution, and
\be
\mbox{}\;\,\,{\cal N}_q^T(x) = N_q^T x^{\alpha_q}(1-x)^{\beta_q}\,
\frac{(\alpha_q+\beta_q)^{(\alpha_q+\beta_q)}}{\alpha_q^{\alpha_q}\beta_q^{\beta_q}}
\,\,\,\,\,\,{\cal N}_q^C(z) = N_q^C z^{\gamma_q}(1-z)^{\delta_q}
\frac{(\gamma_q+\delta_q)^{(\gamma_q+\delta_q)}}{\gamma_q^{\gamma_q}\delta_q^{\delta_q}}\,,
\ee
with $ |N_q^{T(C)}|\leq 1$. With these choices, both $\Delta_T q$ and $\Delta^N\!D_{\pi/q^\uparrow}$ obey their proper positivity bounds. The term $[f_{q/p}(x) +\Delta q(x)]$ in Eq.~(\ref{detq}) is evolved in $Q^2$ using the transversity evolution kernel. Similarly, for the $Q^2$ evolution of the Collins function, which remains so far unknown, we use the unpolarized DGLAP evolution of its collinear factor $D_{\pi/q}(z)$.

We also adopt some additional assumptions: $i)$ for $\Delta_Tq$ we consider only $u$ and $d$ quark contributions, $\alpha_{u,d}=\alpha$, $\beta_{u,d}=\beta$, and $\langle k_\perp^2\rangle_T = \langle k_\perp^2\rangle$; $ii)$ for the FFs we
consider two different expressions for ${\cal N}_q^C$, corresponding
to the so-called ``favoured" ($u\to\pi^+$) and ``unfavoured" ($d\to\pi^+$) FFs,
${\cal N}_{\rm fav}^C(z)$ and ${\cal N}_{\rm unf}^C(z)$, and flavour-independent $\gamma$ and $\delta$ parameters.
We are then left with a total of 9 free parameters:
\begin{equation}
N^T_{u},\, N^T_{d},\, N^C_{\rm fav},\, N^C_{\rm unf},\,
\alpha,\, \beta,\, \gamma,\, \delta, M_h \,.
\label{eq:9-par}
\end{equation}

{}From the fits \cite{Anselmino:2007fs,Anselmino:2008jk} it is clear that SIDIS data
are not presently able to constrain the large $x$ behaviour of the quark
transversity distributions, leaving almost undetermined the values of the parameter $\beta$. This is due to the limited range of Bjorken $x\;(\le 0.3)$ currently explored by HERMES and COMPASS experiments.
This uncertainty has relevant consequences in the study of $A_N$, since its largest values have been measured at large Feynman $x$ values, $x_F \ge 0.3$, and this implies $x\ge 0.3$.

We then devise the following strategy (``scan procedure") to explore the large-$x$ behaviour of $\Delta_Tq$:
$i)$~we perform a 9-parameter ``reference fit" to SIDIS and $e^+e^-$ data taking, w.r.t.~Eq.~(\ref{eq:9-par}), $\beta_u \neq \beta_d$ and $\delta=0$ (this value is indeed preferred by the fit);
$ii$)~we fix the two parameters $\beta_u,\beta_d$ independently in the range $0.0$---$4.0$ by discrete steps of $0.5$ and for each of the 81 pairs of $\beta$s we perform a new 7-parameter fit to SIDIS and $e^+e^-$ data;
$iii)$~we select only those sets of parameters from the scan procedure leading to
a total $\chi^2\le\chi^2_0+\Delta\chi^2$, where $\chi^2_0$ refers to the 9-parameter fit and $\Delta\chi^2$ is the same as that used in Refs.~\cite{Anselmino:2007fs,Anselmino:2008jk} to generate the error band. Notice that all 81 points of our grid in $(\beta_u,\beta_d)$ lead to acceptable fits.
$iv)$~For each set, we calculate the Collins pion SSA in $pp$ collisions in the kinematical
regions of the available data for the E704, STAR (for $\pi^0$) and BRAHMS (for $\pi^\pm$) experiments.
$v)$ Finally we generate a ``scan band", by taking the envelope of ALL curves for $A_N(\pi)$ obtained from the scan procedure. This band shows the potential role of the Collins effect alone in describing $A_N(p^\uparrow p\to\pi X)$ data while preserving a combined fair description (quantified by $\Delta\chi^2)$ of the SIDIS and $e^+e^-$ data on Collins azimuthal asymmetries.

\subsection{Results}
Here we present a selection of results obtained with the Kretzer set and the
unpolarized-like Collins evolution.\footnote{The use of a different FF set and/or of a transversity-like Collins evolution does not lead to any significant difference in our conclusions.}

\begin{figure}
\begin{minipage}[c]{19cm}
\includegraphics[width=7cm,angle=0]{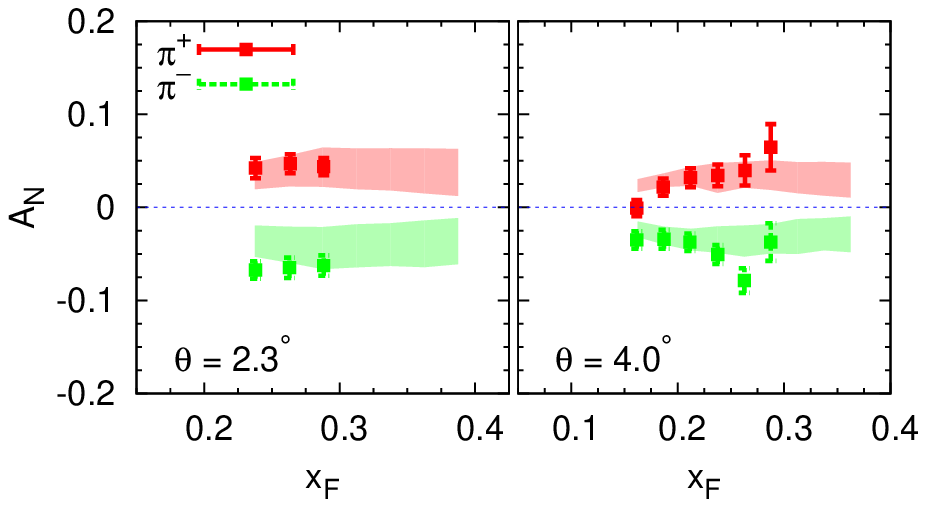}
\includegraphics[width=7cm,angle=0]{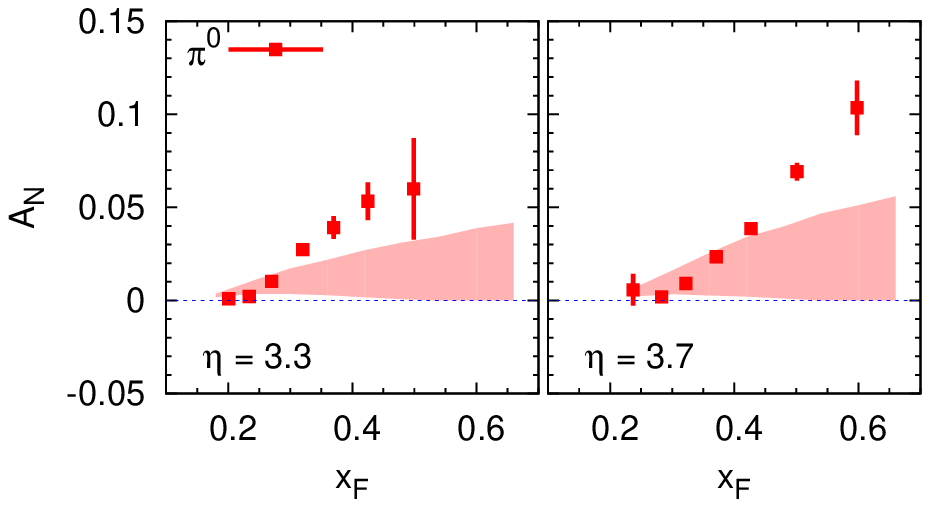}
\end{minipage}
\caption{
Left panel: scan band for the Collins contribution to $A_N(\pi^\pm)$, as a
function of $x_F$ at two different scattering angles, compared with BRAHMS data \cite{Lee:2007zzh}.
The shaded band is generated following the procedure explained in the text. Right panel: same for the Collins contribution to $A_N(\pi^0)$, as a function of $x_F$ at two different rapidity values, compared with STAR data \cite{Abelev:2008af}.}
\label{fig:an-brahms}
\end{figure}

In Fig.~1, left panel, the scan band for $A_N$ is shown for charged pions and
BRAHMS kinematics, while in the right panel the same result is given for neutral pions and STAR kinematics. These results give already some clear indications:
$i)$ the Collins contribution to $A_N$ is not suppressed as claimed in
Ref.~\cite{Anselmino:2004ky};
$ii)$ alone it might be able to explain the BRAHMS results on $A_N$ for $\pi^\pm$;
$iii)$ $\pi^0$ STAR data on $A_N$ cannot be explained, being the Collins contribution not sufficient for the medium-large $x_F$ SSA data, $x_F \ge 0.3$.

To fully assess the role of the Collins effect in $A_N$ for $\pi^0$ at large $x_F$ for the STAR kinematics, we then perform several further tests.
Firstly we consider explicitly each of the curves from the scan, isolating the set leading to the largest $A_N$ in the large $x_F$ region and then evaluate the corresponding statistical error band. Our result is presented in Fig.~2, left panel. Again, it appears that the Collins effect alone cannot account for the large $x_F$ data. Secondly we repeat our scan procedure by starting from a preliminary reference fit with 13 free parameters (i.e.~allowing for a much larger flavour dependence),
\begin{equation}
N_u^T,\,N_d^T,\,\alpha_u,\,\alpha_d,\,\beta_u,\,\beta_d,\,
N_{\rm fav}^C,\,N_{\rm unf}^C,\,
\gamma_{\rm fav},\,\gamma_{\rm unf},\,\delta_{\rm fav},\,
\delta_{\rm unf},\,M_h\,.
\label{eq:11-par}
\end{equation}
We generate again the scan band with $\beta_{u,d}$ fixed (via 11 parameter fits) and then compute the statistical error band of the ``optimal" set in the scan (see Fig.~2, right panel). Also in this case we cannot describe the full amount of the $\pi^0$ STAR data.

Finally, if applied to the E704 results, the scan band, almost sufficient for the neutral pion SSA, misses completely the $A_N$ data for charged pions at large $x_F$.

\begin{figure}
\begin{minipage}[c]{19cm}
\includegraphics[width=7cm,angle=0]{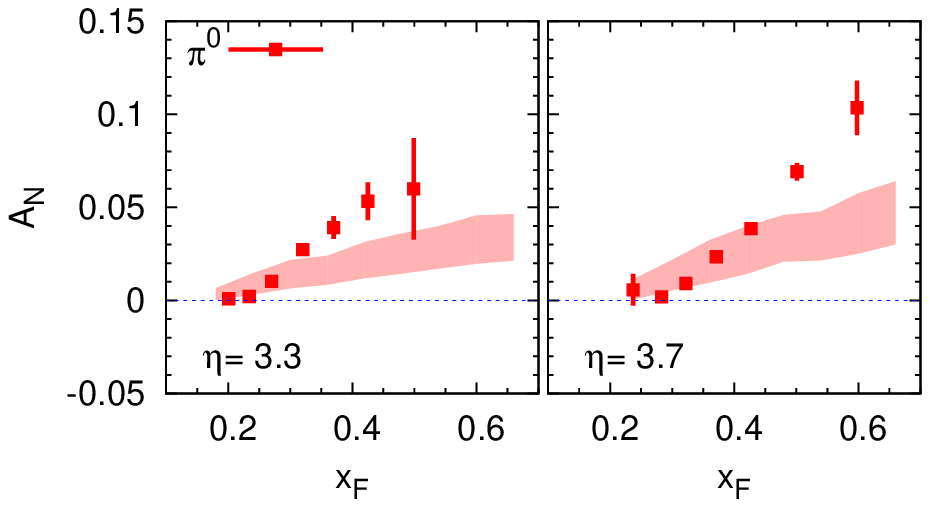}
\includegraphics[width=7cm,angle=0]{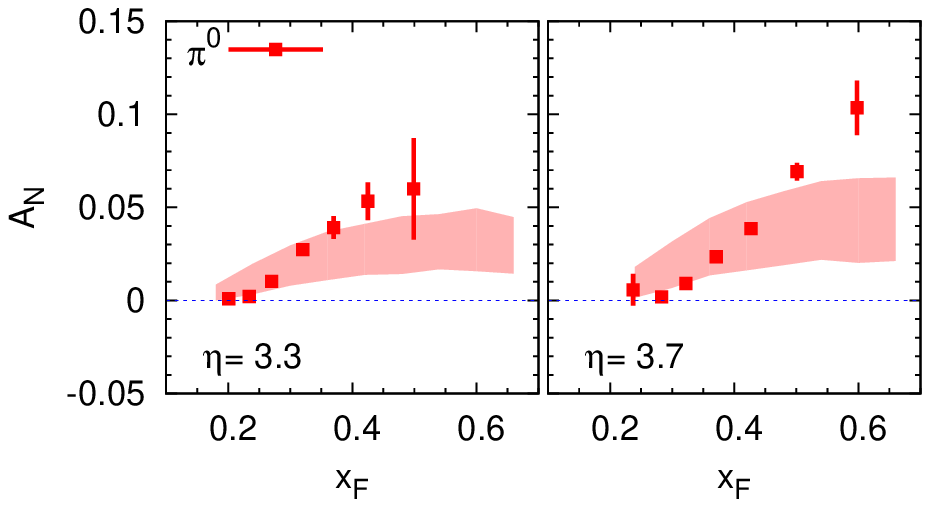}
\end{minipage}
\caption{
Left panel: The Collins contribution to $A_N(\pi^0)$, compared with STAR data at two fixed pion
rapidities \cite{Abelev:2008af}. The shaded band is the statistical error band
generated starting from the 7-parameter optimal fit in the grid procedure. Right panel: same but this time with the statistical error band generated starting from the scan procedure with
11 free parameters.}
\label{fig:an-star-free7}
\end{figure}

Summarizing, based on SIDIS and $e^+e^-$ data, the Collins effect alone seems to be able to reproduce the available data on pion SSAs, only in the region $x_F \le 0.3$.
Above that, the Collins effect is not sufficient and additional mechanisms are required, like, for instance, the Sivers effect. A phenomenological study along this line is underway.

\end{document}